\documentclass[conference,compsoc]{IEEEtran}

\ifCLASSOPTIONcompsoc
\usepackage[nocompress]{cite}
\else
\usepackage{cite}
\fi

\usepackage{amsmath,amssymb,amsfonts}
\usepackage{algorithmic}
\usepackage{graphicx}
\usepackage{textcomp}
\usepackage{xcolor}

\usepackage{hyperref}
\usepackage{subcaption}
\usepackage{booktabs}
\usepackage{listings}

\usepackage{xcolor}
\definecolor{TEETrusted}{HTML}{FFF2CC}

\renewcommand{\paragraph}[1]{\textbf{#1.}}

\usepackage{tikz}
\usetikzlibrary{shapes.geometric}
\usetikzlibrary{arrows.meta, arrows}
\tikzset{
    font=\sffamily,
    x = 1em,
    y = 1em,
    outer sep = 0
}

\usepackage{algorithm}

\usepackage{amsthm}

\theoremstyle{plain}
\newtheorem*{prop*}{Property}
\newtheorem{prop}{Property}

\theoremstyle{definition}
\newtheorem{threat}{Threat}


\theoremstyle{definition}
\newtheorem{obj}{Objective}


\newcommand{\Manu}{\ensuremath{\mathcal{M}}}
\newcommand{\PSvc}{\ensuremath{\mathcal{P}}}
\newcommand{\POwn}{\ensuremath{\mathcal{O}}}
\newcommand{\ADev}{\ensuremath{\mathcal{D}}}
\newcommand{\AUsr}{\ensuremath{\mathcal{U}}}
\newcommand{\BDev}[1]{\ensuremath{D_{#1}}}
\newcommand{\Prog}{\ensuremath{P}}

\newcommand{\Hash}{\ensuremath{ \mathsf{Hash} }}

\newcommand{\kpub}{\ensuremath{ k_\mathsf{pub} }}
\newcommand{\kprv}{\ensuremath{ k_\mathsf{prv} }}

\usepackage[printonlyreused]{acronym}

\newacro{AES}[AES]{{\em Advanced Encryption Standard}}
\newacro{ATF}[ATF]{{\em ARM Trusted Firmware}}
\newacro{CSU}[CSU]{{\em configuration security unit}}
\newacro{DMA}[DMA]{direct memory access}
\newacro{FPGA}[FPGA]{{\em field-programmable gate array}}
\newacro{FSBL}[FSBL]{{\em first-stage bootloader}}
\newacro{MAC}[MAC]{{\em message authentication code}}
\newacro{MMIO}[MMIO]{memory-mapped I/O}
\newacro{OCM}[OCM]{on-chip memory}
\newacro{OTP}[OTP]{{\em one-time-programmable}}
\newacro{PKI}[PKI]{{\em public key infrastructure}}
\newacro{PL}[PL]{{\em programmable logic}}
\newacro{PMU}[PMU]{{\em Platform Management Unit}}
\newacro{PS}[PS]{{\em processing system}}
\newacro{ROM}[ROM]{read-only memory}
\newacro{RSA}[RSA]{Rivest-Shamir-Adleman}
\newacro{SHA}[SHA]{{\em secure hash algorithm}}
\newacro{SoC}[SoC]{{\em system-on-a-chip}}
\newacro{TEE}[TEE]{{\em trusted execution environment}}
\newacro{TCB}[TCB]{trusted computing base}

\begin{document}

\newcommand{\THIS}{T-Edge}
\title{\THIS: Trusted Heterogeneous Edge Computing}

\author{
  \IEEEauthorblockN{
    Jiamin Shen,
    Yao Chen,
    Weng-Fai Wong,
    Ee-Chien Chang
  }
  \IEEEauthorblockA{
    \textit{School of Computing} \\
    \textit{National University of Singapore} \\
    \{shen\_jiamin,changec\}@comp.nus.edu.sg, \{yaochen,wongwf\}@nus.edu.sg
  }
  \IEEEcompsocitemizethanks{
    \IEEEcompsocthanksitem
    This paper has been accepted for publication in Annual Computer Security Applications Conference (ACSAC '24).
    The final version of this paper is available at: \url{https://doi.org/10.1109/ACSAC63791.2024.00027}.
  }
}

\maketitle

\begin{abstract}
  Heterogeneous computing, which incorporates GPUs, NPUs, and FPGAs, is increasingly adopted to improve the efficiency of computer systems.
  However, this shift has given rise to significant security and privacy concerns, especially when the execution platform is remote.
  One way to tackle these challenges is to establish a trusted and isolated environment for remote program execution while maintaining minimal overhead and flexibility.
  While CPU-based trusted execution has been extensively explored and has found commercial success, extension to heterogeneous computing systems remains a challenge.
  This paper proposes a practical trusted execution environment design for ARM/FPGA System-on-Chip platforms, leveraging TrustZone's unique characteristics.
  The design features a dedicated security controller within the ARM TrustZone, overseeing FPGA reconfiguration and managing communication between CPU cores and FPGA fabrics.
  This design involves a provisioning service (\PSvc{}) that enables application users (\AUsr{}) to establish trust in the FPGA fabric within cloud-based computing resources provided by the platform owner (\POwn{}), running applications developed by third-party developers (\ADev{}) and hardware manufactured by the device manufacturer (\Manu{}).
  To ensure the security of our proposed system, we employ an automated protocol verifier, ProVerif, to validate its compliance with essential security requirements.
  Furthermore, we demonstrate the practicality of our system model by implementing a prototype application on the Xilinx MPSoC development board.
\end{abstract}


\section{Introduction}

The increasing demand for processing data in real-time, coupled with the widespread use of Internet of Things (IoT) devices, has prompted a significant shift in computing practices towards what is commonly referred to as the ``edge''.
This includes higher-end IoT devices, such as smart meters and cameras, as well as boundary network devices when adopting a broader definition of edge computing.
This transition has been motivated by the aim of reducing response times, minimizing delays, and conserving network bandwidth by, processing data as close as possible to its origin.
In pursuit of greater efficiency in edge computing, there has been a growing adoption of hardware acceleration solutions to improve the performance of computational tasks.
Among these hardware accelerators, \ac{FPGA} stands out due to its distinctive features, such as reconfigurability, natural parallelism, and direct access to I/O interfaces, rendering it an ideal choice in such scenarios.

\begin{figure}
	\centering
	\includegraphics[scale=0.6]{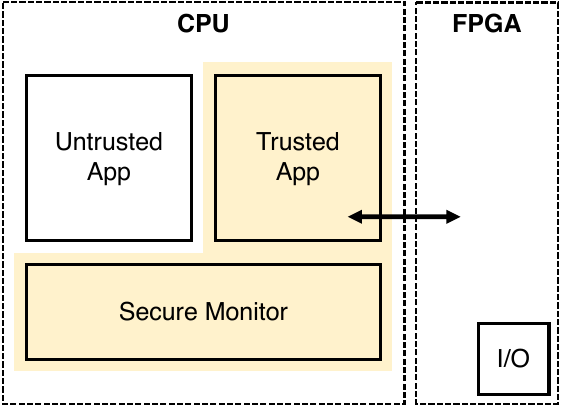}
	\caption{Pairing CPU-based TEE with FPGA.
		The FPGA fabric needs to be involved in the trust boundary.
		It's essential to secure the link between the CPU and FPGA.}
	\label{fig:pairing}
\end{figure}

In the context of edge computing systems, where deployments often occur in untrusted environments with the possibility of physical access by attackers, security concerns have been a significant focus.
Adversaries may load arbitrary software components or hardware configurations onto devices to compromise the system or leak sensitive information.
To address this, \ac{TEE} is required on remote machines to secure and attest applications, even under adversarial physical access.
While existing \ac{TEE} solutions, such as Intel Software Guard Extensions (SGX) or ARM TrustZone, are available for general-purpose CPUs, similar solutions for hardware accelerators like FPGAs are lacking, possibly due to constraints in integrating necessary mechanisms such as attestation and secure boot.
Instead of building the trusted environment directly on FPGAs, one alternative option is to pair a CPU-based TEE with the FPGA, using a software solution to extend the trust, as illustrated in \autoref{fig:pairing}.

However, there are several potential attack vectors that must be considered in this approach:
(1) Given that \ac{FPGA} can be reconfigured for different tasks, a local attacker may manipulate the \ac{FPGA} by loading a malicious bitstream, which could result in incorrect outputs or the leakage of sensitive information.
(2) A malicious \ac{FPGA} configuration has the potential to compromise the memory isolation enforced by the TEE, enabling unauthorized data transfers across partitions \cite{ches/JacobHZRS17,socc/Benhani0AB17}.
(3) The physical link connecting the CPU and \ac{FPGA} is susceptible to physical attacks, such as probing and tampering, which could lead to data leakage and unauthorized access during data transfer.

Addressing these challenges using the paired CPU-based TEEs is not straightforward.
To mitigate (1) and (2), a mechanism is needed to authenticate the \ac{FPGA} configuration without sacrificing the flexibility of FPGA application development and deployment, while providing remote users with proof of authentication.
Regarding (3), secure communication between the CPU and FPGA must be established through a channel isolated from the external environment.
While cryptographic primitives can serve this purpose, protecting key materials on the FPGA side from tampering or exfiltration by a potentially malicious FPGA configuration is not straightforward amidst other design considerations.

In this paper, we present a neat and practical solution to extend the environment from CPU-based TEEs to FPGAs.
Our approach targets the ARM/FPGA \ac{SoC} platform, where ARM cores and \ac{FPGA} fabrics coexist on a single chip.
This approach leverages readily available off-the-shelf ARM/FPGA cards like Xilinx Zynq \cite{xilinx:zynq-7000}, Zynq MPSoC \cite{xilinx:zynqmp-trm}, and Altera Cyclone V SoC FPGA \cite{cyclone}.
The close composition of ARM cores and FPGA fabrics within a single chip elegantly resolves the problem of secure communication between the CPU and FPGA.
Since both components reside on the same chip, the connection between them remains isolated from external exposure, inherently safeguarding against potential physical link vulnerabilities.
There's also an alternative solution from CISCO \cite{cisco/ncs540}, which addresses a similar problem in a network processing scenario.
We discuss it further in \autoref{sec:related-works}.

With such an architecture in place, our design capitalizes on ARM TrustZone due to its clear separation of two ``worlds''.
Intel SGX is not suitable for our scenario for several reasons, despite its flexibility in supporting multiple ``enclaves''.
The flexibility offered by SGX is unnecessary for our application and, more critically, it introduces various side channels.
\autoref{sec:cpu-tee} provides a more detailed discussion on this comparison.

However, the clear separation provided by ARM TrustZone leads to difficulty for third-party developers wishing to develop their own applications.
The prevailing practice for ARM TrustZone mandates that trusted applications be released together with the operating system.
The entire system, including the trusted firmware, needs to be compiled into a single image and signed in its entirety for verification.
This approach restricts developer flexibility, as it requires the entire system to be rebuilt and necessitates a single trusted vendor, typically the OEM, to compile and sign the image.
To address this limitation, an attestation protocol is necessary to shift from traditional verified booting,
where the boot process aborts if signature verification fails, to attested booting,
where the boot process continues and verification happens later.
The protocol should separately handle the verification of the firmware and the attestation of the trusted operating system running in the secure world.
This approach allows third-party developers to create their own trusted applications
and enables remote users to make informed decisions about trusting the system.

In this paper, we focus on tackling the challenges of
(1) authenticating the FPGA configuration,
(2) establishing trust with remote users, and
(3) supporting third-party developers.
We also utilized an automatic protocol verifier, ProVerif \cite{ProVerif}, to confirm adherence to the essential security requirements.
To further substantiate the feasibility of our system model, we have implemented a prototype based on the Xilinx ZCU106 ARM/FPGA development board.
This practical demonstration underscores the feasibility of our proposed methodology.

In summary, this paper presents the following contributions:
\begin{itemize}
	\item Identification of the vulnerability of the security communication between the CPU and FPGA to physical attacks, with the observation that ARM/FPGA \ac{SoC} effectively addresses this issue.
	\item Proposal of a systematic design for a \ac{TEE} tailored to ARM/FPGA SoC, along with an attestation protocol enabling the platform to demonstrate system integrity and authenticity.
	\item Development of a proof-of-concept implementation of the proposed system, including the essential infrastructure required to support the attestation service.
\end{itemize}

In summary, this paper is organized as follows.
We begin with \autoref{sec:preliminaries}, where we introduce some basic concepts.
In \autoref{sec:problem}, we formulate the problem addressed in this paper.
The design of our system is explained in \autoref{sec:system-design}, and we analyse its security properties in \autoref{sec:security-analysis}.
We then illustrate how the design benefits a real-world application with an example in \autoref{sec:case-study}.
\autoref{sec:discussion} discusses decisions that impact the design, while related works are explored in \autoref{sec:related-works}.
Finally, we conclude the paper in \autoref{sec:conclusion}.

\section{Preliminaries}\label{sec:preliminaries}

\subsection{Trusted Execution Environment (TEE)}

Establishing trust over the internet presents significant challenges due to inherent security risks, including identity confirmation and the potential compromise or impersonation of devices.
This poses a severe danger when transmitting sensitive information.
Trusted computing technology aims to address these concerns by creating a secure and trustworthy computing environment.
It employs hardware and software mechanisms to provide a secure platform for executing and storing sensitive data and code.

The objectives of trusted computing are twofold.
Firstly, it aims to ensure that a computing system behaves in a predictable and verifiable way.
This involves creating an execution environment that cannot be compromised by an adversary.
Secondly, trusted computing seeks to convince remote users that legitimate software is running in a trustworthy and secure environment.
This is achieved by establishing a secure channel for communication if necessary, considering that adversaries may have varying attack capabilities, ranging from local adversaries with physical access to the computation device to remote adversaries who can only access the environment remotely.

To achieve these goals, trusted computing employs various techniques.
These include secure storage to guarantee the integrity and confidentiality of the root of trust, secure boot to ensure the system boots from a trusted software image and prevents the execution of malicious or unverified code during the boot process, and essential cryptography primitives such as encryption and digital signatures.
Additionally, additional security measures are often necessary to isolate the \ac{TEE} from the rest of the system, mitigating the risk of adversary compromise.
Collectively, these techniques create a dependable and trustworthy execution environment that is resilient against a broad spectrum of attacks targeting the confidentiality, integrity, and availability of data and code.

\subsection{Existing CPU-based TEE}\label{sec:cpu-tee}

\begin{figure}
    \begin{subfigure}[b]{0.48\linewidth}
        \centering
        \includegraphics[scale=0.64]{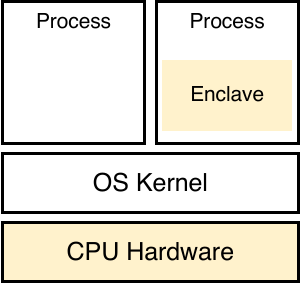}
        \caption{Intel SGX}\label{fig:archsgx}
    \end{subfigure}
    \begin{subfigure}[b]{0.48\linewidth}
        \centering
        \includegraphics[scale=0.64]{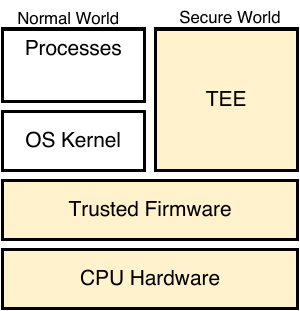}
        \caption{ARM TrustZone}\label{fig:archtz}
    \end{subfigure}
    \caption{
        Comparison between the architectures of Intel SGX and ARM TrustZone.
        The \colorbox{TEETrusted}{shaded} blocks represent the trusted components.
    }
\end{figure}

Existing trusted computing and secure hardware platforms include Intel Software Guard Extensions (Intel SGX) \cite{isca/McKeenABRSSS13,iacr/CostanD16} and ARM TrustZone \cite{trustzone,coinco/NgabonzizaMBCM16}.
These platforms aim to provide secure and trustworthy computation environments for various applications.

Intel SGX (\autoref{fig:archsgx}) offers a hardware-based \ac{TEE} through an instruction set extension, allowing for the creation of secure enclaves that isolate application execution from the rest of the system, thereby protecting against unauthorized access and tampering.
Despite its advantages, SGX's emphasis on multi-tenancy introduces security challenges.
Allowing multiple user enclaves to operate simultaneously necessitates the sharing of resources with potentially untrusted tenants, thereby increasing the system's susceptibility to side-channel attacks\cite{eurosec/GotzfriedESM17}, including those that exploit cache and branch prediction mechanisms.
Additionally, SGX's design, which restricts access to low-level ports and physical memory, presents challenges in efficiently handling I/O-intensive workloads.
Furthermore, the architecture primarily aims to shield user-level applications from the operating system by keeping the TEE at a user-level privilege, which consequently limits its ability to securely manage hardware resources.

Conversely, ARM TrustZone (\autoref{fig:archtz}) offers a reliable solution for trusted computing through the enforcement of a hardware-enforced segregation between secure world and non-secure world (also known as normal world), safeguarding confidential data and code.
It allows the secure world to operate at a higher privilege level, thereby granting exclusive access to essential hardware resources.
However, TrustZone is not devoid of its own set of challenges.
The absence of a standardized isolation mechanism within its secure domain introduces a systemic risk --- a single compromised application could potentially endanger the integrity of the entire system \cite{sp/Cerdeira0FP20}.
This vulnerability necessitates rigorous secure operating system design and implementation.
This thereby complicates deployment flexibility, as the secure OS image, typically integrated with the firmware, requires OEM collaboration for updates.

Furthermore, these platforms are not equipped to manage computation and data-intensive tasks which require the utilization of specialized hardware accelerators such as GPUs, TPUs, or FPGAs.
The situation calls for a novel TEE design that integrates these hardware accelerators to cater to the needs of trusted heterogeneous computing.
To rectify these issues, our proposed solution combines the capabilities of ARM TrustZone and FPGA technology, forming a single-tenant architecture that enhances security and deployment flexibility for a trusted heterogeneous computing platform.

\subsection{Field-Programmable Gate Array (FPGA)}

\ac{FPGA} \cite{sp/Sueyoshi18} is a programmable integrated circuit that can implement user-defined logic circuits.
It is made up of a set of programmable logic blocks, which implements various logic circuits, and I/O ports, which handles the communication with elements outside the FPGA.
There are also wiring elements, which is configurable and connects the components.
With such flexibility, it can be reconfigured on-site to perform a variety of tasks at nearly hardware-level speed.

The configuration of components on the chip add up to a bitstream, essentially serving as a program that runs on the FPGA.
The programming of the bitstream is managed by an embedded processor integrated into the FPGA fabric.
This processor has the responsibility of loading the bitstream, whether from JTAG, other interfaces, or an external storage device, into the FPGA chip.
Additionally, the processor can facilitate the configuration of the FPGA fabric at runtime, a capability known as partial reconfiguration.

\subsubsection{Existing FPGA Security Enforcement}

The existing security enforcement is designed to protect the intellectual property of the hardware design (the bitstream, hereinafter referred to as IP) \cite{csur/TuranV21}.
Thus, the security model consists of two entities, the IP developer and its consumer.
An IP developer develops an IP and sells it to consumers for financial interest, but he is worried that it could be copied and deployed to FPGA instances more than authorized.
Moreover, he is also concerned about the IP being reverse-engineered, leading to a leakage of the hardware design.
If that happens, copycats may use his design to develop other IPs and seize the wealth.
So the IP distributed must be sure that it can only be deployed to a specific FPGA instance, and the consumers cannot learn anything about the hardware design of the IP.

To enforce the security model, FPGA is equipped with \ac{OTP} memory, which can be implemented using eFuse.
The IP developer can install a secret key into the device by programming the \ac{OTP} memory and distributing his IP encrypted using the key.
As the key is known only to the FPGA device and the developer, it prevents malicious consumers from stealing the IP.
The security model works in classic scenarios where the developers distribute the FPGA hardware and the bound program together as a package, but it is not the case in cloud computing.
In our cloud computing setting, the FPGA hardware is owned and provided by cloud computing providers, and IP developers may not program or access the key of the hardware.
Besides, he may not even know which specific FPGA board could be used for consumers computation.

Another inappropriate point is that only the IP developers interest is concerned in this security model, and the security of consumer's data is not considered.
It worked because the hardware is designed to work on-premise close to the consumer.
Thus, the consumer can secure the path between his data and the FPGA device.
However, when the FPGA device is deployed on the cloud, the consumer has no way to build a secure channel to the FPGA hardware without further change.
Therefore, the security model does not fit in our setting.

\subsubsection{ARM/FPGA SoC}

ARM/FPGA \ac{SoC} is a type of \ac{SoC} that integrates FPGA fabric and ARM processors on a single chip, where the FPGA fabric is referred to as \ac{PL} and other components including ARM processors are referred to as \ac{PS}.
The architecture was designed to provide more flexible on-chip control to FPGA platform, and offload part of tasks on ARM processors to programmable hardware.
There are off-the-shelf products implementing this architecture, such as Xilinx Zynq \cite{xilinx:zynq-7000}, Zynq MPSoC \cite{xilinx:zynqmp-trm}, and Altera Cyclone V SoC FPGA \cite{cyclone}.

\subsection{Boot Process and Secure Boot}

\begin{figure}
    \centering
    \includegraphics[scale=0.6]{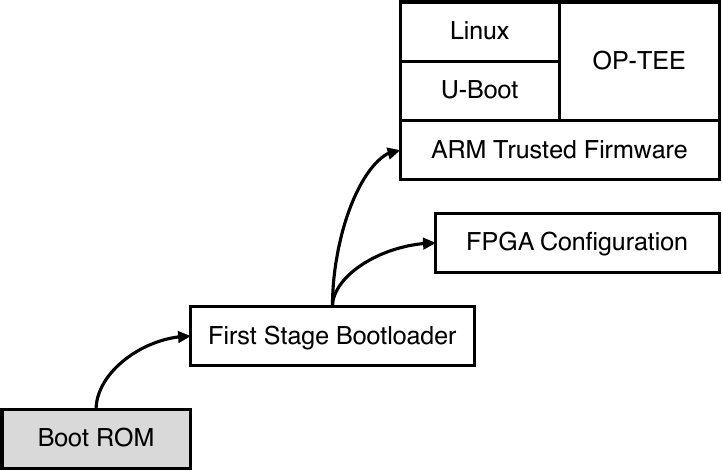}
    \caption{Simplified Boot Sequence of the System. The Boot ROM is an unmodifiable piece of code embedded in the chip.}
    \label{fig:bootseq}
\end{figure}

A simplified boot sequence is illustrated in \autoref{fig:bootseq} to facilitate the description.
Following a system reset, an unmodifiable Boot ROM is executed, initiating the loading and execution of the \ac{FSBL}.
The \ac{FSBL} is responsible for part of hardware initialization and configuration, as well as loading other components of the system.

Secure Boot is a security mechanism that ensures that the system boots using only authorized images.
It is a chained process in which the previous stage's software is responsible for authenticating the next stage's software.
It requires that the images to be loaded in each stage of the boot process be cryptographically signed using a private key.
The corresponding public key, or a hash of it, is stored immutably in the hardware.
The public key is used to verify the signature of the image before it is loaded and executed.

The boot process starts with the Boot ROM, which serves as the platform's immutable root of trust.
Since the Boot ROM is hardcoded into the hardware, it cannot be modified.
It is responsible for loading the \ac{FSBL}, the first configurable software component, from flash memory and verifying its authenticity.
The FSBL, when being released, is signed using the vendor's private key, with the corresponding public key and signature attached to the FSBL image.
During provisioning, the hash of the public key is embedded into the device.
The Boot ROM first checks that the public key matches the stored hash, then uses the public key to verify the FSBL's signature.

\section{Framework Formulation}\label{sec:problem}

We propose a systematic design specifically targeting ARM/FPGA SoC platforms that pairs a CPU-based \ac{TEE} with \ac{FPGA} fabric.
This design includes a complete ecosystem that allows developers to flexibly develop applications, alongside an architecture and attestation protocol that enforce system security.

In this section, we describe the system model in which our proposed method operates, outline the threat model, and  define the security objectives the system aims to achieve.

\subsection{System Model}\label{sec:system-model}

\begin{table}
	\centering
	\caption{System Model Parties}
	\label{tab:parties}
	\begin{tabular}{cc}
		\toprule
		\textbf{Code} & \textbf{Description}  \\
		\midrule
		\Manu{}       & Device Manufacturer   \\
		\PSvc{}       & Provisioning Service  \\
		\POwn{}       & Platform Owner        \\
		\ADev{}       & Application Developer \\
		\AUsr{}       & Application User      \\
		\bottomrule
	\end{tabular}
\end{table}

\begin{figure}
	\centering
	\includegraphics[scale=0.64]{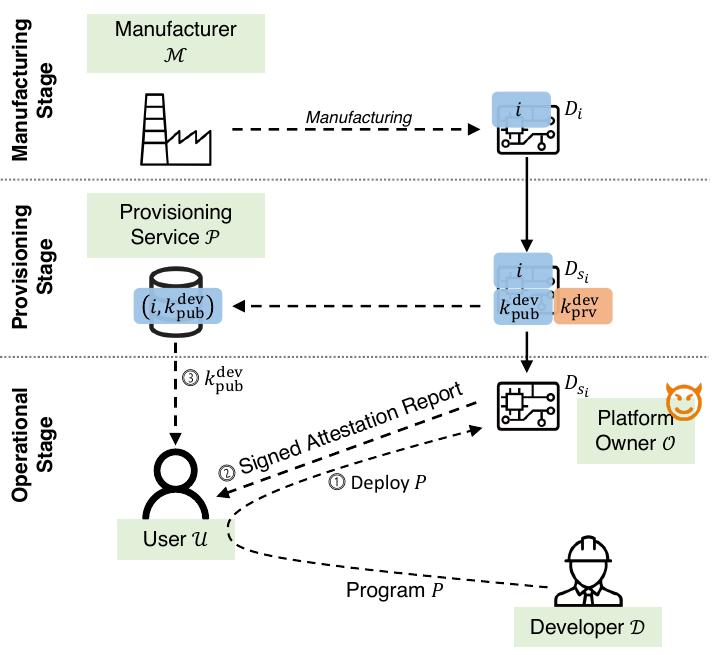}
	\caption{System Model}
	\label{fig:system-model}
\end{figure}

We address the problem within the system model depicted in \autoref{fig:system-model}, which includes five participants outlined in \autoref{tab:parties}.
From a temporal perspective, a device's life cycle can be divided into three stages: the manufacture stage, the provisioning stage, and the operational stage.

\paragraph{Manufacture Stage}
During the hardware manufacture stage, the device manufacturer \Manu{} produces the device and assigns a distinct, non-modifiable serial number to it.
This serial number is publicly visible and is used to reference the device.
To reference a device with a serial number of $i$, we use the notation \BDev{i}.
Following the manufacture stage, \Manu{} hands over the device to a provisioning service provider \PSvc{} for initial secret provisioning.

\paragraph{Provisioning Stage}
In the provisioning stage, the provisioning service provider \PSvc{} executes a provisioning bootloader on the device to generate a device private key $\kprv^{\text{dev}}$.
This private key is then burned into the device's \ac{ROM} and never leaves the device.
The corresponding device public key $\kpub^{\text{dev}}$ is output by the bootloader.
\PSvc{} then publishes the mapping between $i$ and $\kpub^{\text{dev}}$.
For a device to claim it is $\BDev{i}$, it must prove to the verifier that it possesses the private key corresponding to $\kpub^{\text{dev}}$.
\PSvc{} is also responsible for releasing the \ac{FSBL}, which is signed using its public key $\kpub^{\text{boot}}$.
This public key is also burned into the device so that the device can verify the \ac{FSBL} signature.
After the provisioning stage, the device is sold to a platform owner, identified as \POwn{}.

\paragraph{Operational Stage}
During the operational stage, an application, denoted as \Prog{}, is developed by an application developer \ADev{}, and subsequently deployed to some provisioned device \BDev{i}.
During the boot process, an attestation report is generated by \BDev{i} with the device's identity $i$ and the application's identity $\Hash(\Prog{})$ included.
This report is then signed by the device private key $\kprv^{\text{dev}}$.
After the application is up, its user, identified as \AUsr{}, initiates the attestation process to verify that \Prog{} is running on the \BDev{i}.
To verify the attestation report, \AUsr{} must obtain the corresponding public key used to verify the attestation report off-chain from a trusted source.
By verifying the attestation report, \AUsr{} can confirm that \BDev{i} is running the intended \Prog{}.

\subsection{Threat Model}\label{sec:threat-model}

As described in \autoref{tab:parties}, the system model involves five parties, and each of them has a specific role as described in the system model (\autoref{sec:system-model}).

\paragraph{Device Manufacturer (\Manu{})}
\Manu{} is assumed to be a trusted entity with regard to designing and producing the device securely.
It is assumed that the device produced by a trusted \Manu{} will provide proper hardware support for the root-of-trust and secure boot mechanism, and will have the security properties claimed by TrustZone.
However, \Manu{} is motivated to collect information about the applications running on the devices for marketing or profiling purposes.
Such information can be used to infer the user's behaviour or preferences, which may be considered a privacy violation.

\begin{threat}
	\label{threat:manu:privacy}
	\Manu{} may be curious about what application is running on which device.
\end{threat}

\paragraph{Provisioning Service (\PSvc{})}
The pair of $i \mapsto \kpub^{\text{dev}}$ published by the \PSvc{} is assumed to be correct.
\PSvc{} is also assumed to be honest and conduct a comprehensive security review of the \ac{FSBL} prior to its release.
Similar to \Manu{}, \PSvc{} is also motivated to collect usage information.

The \PSvc{} and the \Manu{} could be merged into a single entity without exposing to the threat.
However, we choose to separate them for segregation of duties.
This segregation allows users with partial trust in the \Manu{} to turn to their own trusted \PSvc{}, thus mitigating the risk of a single point of failure in case the \Manu{} is not consistently trusted.

\begin{threat}
	\label{threat:psvc:privacy}
	\PSvc{} may be curious about what application is running on which device.
\end{threat}

\paragraph{Platform Owner (\POwn{})}
The primary security threat originates from the \POwn{}, who has physical control over the device.
The \POwn{} could carry out physical attacks on the device.
Nonetheless, we assume that through physical attacks, the attacker is unable to compromise the chip, such as by probing the internal state or intercepting communication between components on the chip.
Though unable to compromise the chip, \POwn{} could still load arbitrary applications onto the device, intending to deceive the remote user or extract the device's secret through software-based attacks.
Additionally, \POwn{} may attempt to execute legitimate applications outside the trusted environment, thus enabling exfiltration or tampering with the internal state of applications.

\begin{threat}
	\label{threat:pown:prog}
	\POwn{} may load a different application in the TEE.
\end{threat}

\begin{threat}
	\label{threat:pown:fenv}
	\POwn{} may execute legitimate applications outside the TEE.
\end{threat}

\begin{threat}
	\label{threat:mitm}
	\POwn{} may try to intercept the communication between the remote user and the TEE.
\end{threat}

\paragraph{Application Developer (\ADev{})}
We assume that \ADev{} develops applications that are free from vulnerabilities and leak no application data from the user \AUsr{}.
The development of a secure application, for example \cite{acsac/WanS00H20}, is orthogonal to our problem and is out of the scope of this paper.
However, a malicious \ADev{} may deliberately develop and deploy an application that tries to exfiltrate the device's private key to compromise the device.

\begin{threat}
	\label{threat:adev:prog}
	\ADev{} may develop a malicious application that tries to exfiltrate the device's private key.
\end{threat}

Furthermore, cryptographic primitives used in the system are assumed to be sufficiently robust to achieve their objectives despite potential attacks by adversaries.
While TrustZone vulnerabilities may exist, either now or in the future, we assume that these will be patched as they affect the entire range of security applications, including well-established ones.

\subsection{Security Objectives}\label{sec:security-objectives}

Given the threat model and assumptions outlined above, the system provides attestation against the device, and protect user data and results by fulfilling the following security objectives.

\begin{obj}[Fault Isolation]
	\label{obj:fault-isolation}
	Malicious application being executed on a device will not leak $\kprv^{\text{dev}}$.
\end{obj}

This is to prevent \autoref{threat:adev:prog}.
This ensures that the device is still trustworthy even if a malicious application was deployed on it, as long as the current application is legitimate.

\begin{obj}[Device Authentication]
	\label{obj:device-auth}
	Remote users can verify the identity of the device on which a program is running.
\end{obj}

This is to prevent \autoref{threat:pown:fenv}.
It determines whether the program operates on a secure device and, in cases where multiple devices of the same type exist, identifies the intended device.
Device authentication is key to preventing attacks in which adversaries attempt to impersonate legitimate devices so as to carry out unauthorized actions or access sensitive information.

\begin{obj}[Program Authentication]
	\label{obj:prog-auth}
	Remote users can achieve a measurement of the program running on the device.
\end{obj}

This is to prevent \autoref{threat:pown:prog}.
It ensures that application users can authenticate the binary program running on their device and confirm whether it is the intended version or a potentially modified, malicious version.
This authentication is crucial in thwarting attacks where adversaries try to substitute a legitimate program with a malicious one, enabling them to execute unauthorized actions or access sensitive data.

\begin{obj}[Secure Channel]
	\label{obj:secure-channel}
	A secure channel is established between the device and the remote user.
\end{obj}

This is to prevent \autoref{threat:mitm}.
It ensures that the communication between the device and the remote user is confidential and tamper-proof, even in the presence of adversaries.
It is essential to protect the confidentiality and integrity of the data exchanged between the device and the remote user.

\begin{obj}[Privacy]
	\label{obj:attest-privacy}
	The device's identity and the program's identity are protected from the device manufacturer and the provisioning service.
\end{obj}

This is to prevent \autoref{threat:manu:privacy} and \autoref{threat:psvc:privacy}.
It ensures that the \Manu{} and the \PSvc{} cannot determine which program is running on a device.

These security objectives ensure that a high level of trust between the application user and the device is established.

\section{System Design}\label{sec:system-design}

In this section, we present the system design of our \ac{TEE} with \ac{FPGA} computational resources.
The design aims to achieve enhanced performance through FPGA acceleration while providing flexibility by allowing third parties to develop and deploy their applications on the platform.
It also ensures and enforces the security properties that make it a TEE, including proper isolation of secure resources from untrusted programs, secure measurement and attestation of the system to establish trust between the remote user and the TEE, and secure communication between them.

\subsection{System Architecture}

The proposed system is built upon the existing ARM/FPGA \ac{SoC} platform, which has \ac{FPGA} fabric and ARM cores integrated on a single chip with full ARM TrustZone support.
We extend the conventional TrustZone-based \ac{TEE} design, taking advantage of its ability to isolate secure peripherals from untrusted software access.

\subsubsection{Isolated CPU-FPGA Communication}

The communication between the CPU and the \ac{FPGA} can be compromised either by hacking the hardware wiring between the CPU and the FPGA or by attacks from software on the CPU.
Regarding the hardware wiring, the \ac{SoC} platform conceals the physical connection between the CPU and the FPGA inside the chip.
Assuming hacking into the chip is not feasible, the hardware communication between the CPU and the FPGA is inherently secured.
Software communication between the CPU and the FPGA is done through \ac{MMIO}.
This means that software running on the CPU communicates with the FPGA by reading from or writing to segments of physical memory space where the FPGA resources are mapped.
With the help of TrustZone, the CPU's physical memory space is divided into secure and normal worlds, with access control enforced by the hardware.
By mapping the FPGA resources to the secure world, or configuring the memory segments mapped to the FPGA as secure-world-only, the system can prevent untrusted software from accessing FPGA resources.
This setting effectively shields FPGA resources from potential interference by the normal OS, reinforcing the protection and integrity of the secure execution environment.

\subsubsection{Isolated FPGA Configuration}

The very first \ac{FPGA} configuration happens at boot time, with the bitstream loaded into the \acf{PL} by the \ac{FSBL}.
Certain FPGA devices support runtime reconfiguration, enabling the PL to be reconfigured without resetting the entire system.
While this feature allows for fast switching between different applications, it could introduce a potential vulnerability, as an adversary might exploit it to execute a malicious application on the FPGA after the measurement and attestation process is completed.
To address this issue, the architecture shall require that reconfiguration be managed by TrustZone, ensuring that the PL can only be reconfigured by the trusted application running in the secure world.
In this scenario, runtime reconfiguration becomes an application-level behaviour, defined as part of the trusted application's logic.
This ensures that only authorized and verified changes to the FPGA configuration can occur.
If this option is not available on a specific device or is not needed for an application, the runtime reconfiguration capability must be explicitly disabled from the FPGA side in the bitstream.
This effectively prevents the PL from being reconfigured without a complete system reset, thereby preventing unauthorized changes to the FPGA configuration by adversaries without redoing the attestation process.

\subsection{Remote Attestation Protocol}

The remote attestation protocol involves two main phases:
(1) \textbf{At Boot-Time:}
This phase occurs during system boot.
It entails generating measurements of the initial system state and creating an ephemeral key pair.
These measurements are used to establish the integrity of the system, while the key pair is used for subsequent secure communication.
Importantly, this process occurs locally on the device and is not exposed to the external environment.
(2) \textbf{At Run-Time:}
During runtime, an interactive challenge-response protocol takes place between the remote user and the device.
The Secure OS initiates this protocol by presenting an attestation report containing a challenge, signed using the previously generated attestation key pair.
Remote users can verify the authenticity of the attestation key via the provisioning service and then use it to verify the attestation report.
By successfully completing this process and demonstrating possession of the attestation key, the device can assert its identity as the intended device and confirm the execution of the intended application.

\subsubsection{Secure Boot and Measurement}

Traditional secure boot methods require signing entire system images with specific keys.
If the signature verification fails, the device enters a fail-secure state, allowing only authenticated software to run.
However, this approach lacks flexibility and cannot measure the running program.
To provide flexibility for third-party developers, our design limits the use of traditional secure boot to the \ac{FSBL} only.
In our updated protocol, we do not verify the signatures of other boot components.
Instead, we calculate the digests of the initial state of the boot components $\Hash(P)$,
which is the measurement of the initial system state.

During the boot process, as shown in \autoref{fig:bootseq}, the Boot ROM, which is unmodifiable,
is the first piece of code executed after power-on.
The \ac{FSBL} is the first customizable code to run on the system and is responsible for loading other boot components into memory and executing them.
Since malicious programs have no opportunity to execute at that point, the execution of the \ac{FSBL} will not be interfered with.
Thus, we task the \ac{FSBL} with measuring the initial system state.

The \ac{FSBL}, after calculating the measurement, will generate the attestation report, which includes the device's serial number $i$, the measurement of the FSBL $\Hash(F)$, and the measurement of the boot components $\Hash(P)$.
It will also generate a pair of ephemeral keys $\left( \kprv^{\text{s}}, \kpub^{\text{s}} \right)$ for the application's use, and the public key $\kpub^{\text{s}}$ is also included in the attestation report.
The attestation report shall be signed using the device key $\kprv^{\text{dev}}$.
This pair of keys is linked to the device and is used only to sign and verify the attestation report.
After the attestation report is signed, the private key is discarded, so that it is not accessible to the application.
Finally, the attestation report $ \left\langle i, \Hash(F), \Hash(P), \kpub^{\text{s}} \right\rangle $ and the private key $\kprv^{\text{s}}$ are passed to the Secure OS as initial arguments.

\subsubsection{Trust Establishment}

\begin{figure}
    \includegraphics[width=\linewidth]{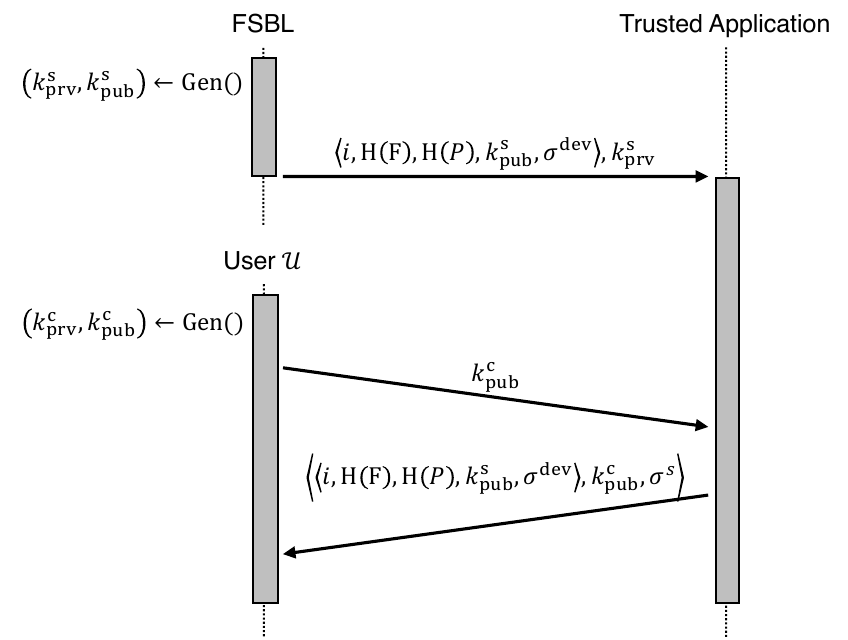}
    \caption{Attestation Protocol}
    \label{fig:protocol}
\end{figure}

The interactive attestation protocol (\autoref{fig:protocol}) follows the challenge-response paradigm to prevent replay attacks.
The challenge is an ephemeral public key $\kpub^{\text{c}}$ generated by the remote user, so that secure channel is established during the attestation process.

At runtime, when a user sends an attestation request with $\kpub^{\text{c}}$ to the device, the trusted application generates a response including the signed attestation report and the challenge, then signs the response using $\kprv^{\text{s}}$.
It then sends the response back to the user.

The user-side procedure involves verifying the attestation report and the challenge response.
First, the remote users retrieve the pair $\left(i, \kpub^{\text{dev}}\right)$ from the provisioning service offline.
Then check the followings:
\begin{enumerate}
    \item The challenge is the same as the one sent by the user.
    \item The challenge response signature is verified using the $\kpub^{\text{s}}$.
    \item The attestation report signature is verified using the $\kpub^{\text{dev}}$.
\end{enumerate}
Upon successful verification of the challenge response and the attestation report, the remote user will have confidence that the application identified by the attestation report is indeed running on that specific device  (\autoref{obj:device-auth}).
The remote user can then decide whether to trust the session based on the measurement $\Hash(F)$ and $\Hash(P)$ attached to the attestation report (\autoref{obj:prog-auth}).

By embedding ephemeral public key into the challenge-response procedure, we can establish a secure channel (\autoref{obj:secure-channel}) between the remote user and the device using Diffie-Hellman key exchange protocol.

\section{Security Analysis}\label{sec:security-analysis}

\newcommand{\FSBL}{\ensuremath{\mathsf{FSBL}}}

In this section, we present an analysis that demonstrates how our proposed system, which operates under the adversary model defined in \autoref{sec:threat-model}, fulfils the objectives defined in \autoref{sec:security-objectives}.

\subsection{Fault Isolation}

In this section, we show how \autoref{obj:fault-isolation} is achieved.

\begin{prop}\label{prop:prvk}
    Applications have no access to the private key $\kprv^{\text{dev}}$ burned into the device.
\end{prop}

The ROM that stores the private key $\kprv^{\text{dev}}$ is only accessible to the \ac{FSBL} during the boot process.
The FSBL reads the private key from the ROM and uses it to sign the attestation report.
Before handing off control to the application, the \ac{FSBL} clears the memory that stores the private key and disables access to the ROM.
Consequently, the only information that the applications have is the signature attached to the signed attestation report and the public key $\kpub^{\text{dev}}$.

According to cryptographic principles, it is infeasible to derive the private key $\kprv^{\text{dev}}$ from the public key $\kpub^{\text{dev}}$ and the signature.
Therefore, the application has no information about the private key $\kprv^{\text{dev}}$, and \autoref{prop:prvk} holds.
Since applications have no access to the private key $\kprv^{\text{dev}}$, malicious applications cannot cause the private key to be leaked, thus achieving \autoref{obj:fault-isolation}.

\subsection{Attestation}

\subsubsection{Device and Program Authentication}

The context of a session that runs \ac{FSBL} $F$ and application $P$ on a device with serial number $i$ can be identified using a tuple
$$
    C = \left\langle i, F, P, \kpub^{\text{dev}}, \kpub^{\text{s}}, \kpub^{\text{c}} \right\rangle
$$
where $\kpub^{\text{s}}$ is the ephemeral public key, and $\kpub^{\text{dev}}$ is the device public key.

\begin{prop}\label{prop:auth}
    For all probabilistic polynomial-time adversary
    without access to a device running under context $C$, where the $P$ in $C$ is a legitimate one,
    and have access to all the history of challenge-response pairs of the device,
    the probability that an adversary succeeds in presenting an attestation response that passes the verification process
    is negligible.
\end{prop}

We first elaborate on the relationship between \autoref{prop:auth} and Security Objectives \ref{obj:device-auth} and \ref{obj:prog-auth}.
Violating \autoref{obj:device-auth} involves (1) executing $P$ on a device with a different serial number, or (2) executing it outside the TEE.
Violating \autoref{obj:prog-auth} involves (3) executing a different program $P'$ on the same device.
In cases (1) and (3), the adversary has access to a device with a different context, and in case (2), the adversary has no access to the device.
Thus, \autoref{prop:auth} implies that \autoref{obj:device-auth} and \autoref{obj:prog-auth} are achieved.

Now we show that \autoref{prop:auth} holds in the proposed system by showing that:
(1) Any valid attestation report must have been generated by $F$ running under context $C$.
(2) Replayed attestation report will be rejected.
Given these two points, the adversary cannot present a valid attestation response that passes the verification process without actually access to the device running under context $C$.

Given \autoref{prop:prvk}, the adversary cannot obtain the private key $\kprv^{\text{dev}}$.
Given the unforgeable property of the signature scheme, the adversary, without knowing $\kprv^{\text{dev}}$, cannot forge a signature and present a valid attestation report.
Thus, an attestation report that passes the signature verification process must have been generated by the \ac{FSBL} $F$.
Since the integrity of the \ac{FSBL} is maintained through the secure boot process, an adversary would not be able to load a malicious \ac{FSBL} onto the system without compromising the secure boot mechanism.
Therefore, $F$ must be legitimate.
Since $F$ is legitimate, it will only sign the attestation report if the device is running under context $C$.
Thus, any valid attestation report must have been generated by the \ac{FSBL} $F$ running under context $C$.

If an attestation report from a previous session with context $C$ is replayed in a new session with context $C'$, the attestation response must (1) contain an old challenge, or (2) contain the new challenge and have the signature generated by the private key $\kprv^{\text{s}}$ in $C$.
For case (1), the attestation response will be rejected because the challenge-response pair has been used before.
For case (2), the adversary having access to $C'$ will not have access to the private key $\kprv^{\text{s}}$ in $C$ since $C$ is a legitimate one, and the signature is unforgeable.
So the adversary cannot present a valid attestation response that passes the verification process by replaying an old attestation report.

Thus, it can be concluded that if an attestation response is successfully verified by the user, it must have been generated by the trusted \ac{FSBL} $F$ running in the legitimate context $C$ in a fresh session.
This satisfies \autoref{prop:auth} and thus the Security Objectives \ref{obj:device-auth} and \ref{obj:prog-auth} are achieved.

\subsubsection{Secure Channel Establishment}

From the above analysis, it can be concluded that the remote user \AUsr{} and the trusted application have common knowledge of their ephemeral public keys $\kpub^{\text{c}}$ and $\kpub^{\text{s}}$.
By running the Diffie-Hellman key exchange protocol, the user and the device can derive a shared secret $k$ from their public keys.
Additionally, since the public keys of the user and the device are both authenticated by the attestation response, there is no way for an adversary to modify the public keys or the signature to tamper with the shared secret.
Thus, the shared secret is guaranteed to be secure and only known to the user and the device, providing confidentiality and integrity protection for the data and computation results.
This achieves \autoref{obj:secure-channel}.

\subsection{Attestation Privacy}

According to the model \autoref{sec:system-model}, the \Manu{} is not involved after the board is sold to the \POwn{}.
It has no knowledge of the subsequent usage of the device, so the \Manu{} cannot link the device to the \AUsr{}, \ADev{}, or the application running on it.
The \PSvc{} is involved in the process only to provide device public keys to the \AUsr{}.
Therefore, it only gains knowledge that some \AUsr{} is interested in some device but does not know the application running on the device.
Thus, \autoref{obj:attest-privacy} is achieved.

\section{Implementation and Evaluation}\label{sec:case-study}

\subsection{Implementation}

To show the effectiveness and efficiency of our system design, we implemented a prototype implementation\footnote{The artifact is available at \url{https://github.com/T-Edge}.} of our system on the Xilinx ZCU106 development board.
The system includes a patched \ac{FSBL} and a patched OP-TEE~\cite{optee}.
The \ac{FSBL} is patched to generate devices key pair using a key derivation function using a symmetric key burned into the device before deployment.
It then captures the measurement of the system image, generates an attestation report, and signs it using the device's private key.
The OP-TEE works as the OS in the secure world, and is patched to support our attestation operations.

When deploying our system on a specific platform, certain configurations must be enforced to ensure security.
For instance, on the ZCU106, JTAG access should be disabled, and secure boot must be enabled by programming the relevant eFUSE for production environments \cite{xilinx:zynqmp-trm}.
Additionally, the ZCU106 has multiple processors on the PS side.
Any unused processor should be configured to operate in the non-secure world to prevent potential attacks.

Although it is outside the scope of this paper, it is worth mentioning that if the FSBL is not properly implemented and the system is booted with a contaminated context, such as maliciously crafted register values or memory contents, the execution of the FSBL could be compromised.
As a general workaround, it is recommended to initialize the context with a known good state at the very beginning of FSBL execution.

\subsection{Case Study}

For the application, we chose a 2D Convolution Filtering Acceleration application~\cite{vitistut}.
In this application, the convolution filter is defined as a 2D matrix.
For each pixel in the image, a pixel set is formed by including the pixel and its neighbouring pixels.
This pixel set is multiplied by the convolution filter and summed up to produce a new pixel value.
The accelerator is implemented in C++ and synthesized into an FPGA IP core using Vitis HLS.
It is important to note that the principles discussed herein are applicable to a broad spectrum of application IP cores, including both complex ones and those involving security considerations.

To assess the impact of attestation on performance, we created two system images: one with attestation and one without.
The version without attestation is based on a standard Linux operating system and includes the accelerator built into the FPGA bitstream, a kernel module for the accelerator, and a corresponding user-space application.
The version with attestation is deployed on the same hardware platform, with the same accelerator, but with the addition of our attestation mechanism.
The accelerator is deployed in the \ac{PL} part, where its interface is mapped to the physical memory space and configured as only accessible from the secure world.

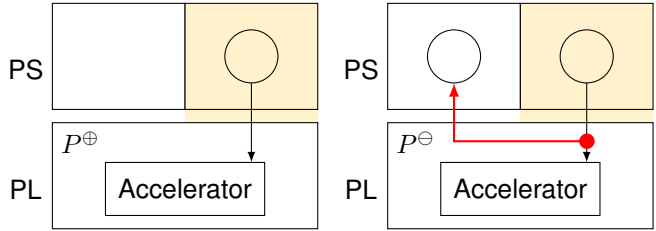
\begin{figure}
	\begin{subfigure}[t]{0.48\linewidth}
		\centering
		\begin{tikzpicture}
    \fill [fill=TEETrusted] (5,3.5) rectangle ++(5,0.5);

    \draw (0,-0.5) rectangle ++(10, 4) coordinate[midway] (c) {};
    \draw (-1,1) node {PL};
    \node [draw, rectangle, minimum width=6em, minimum height=2em] at (5, 1) (d) {Accelerator};
    \node [anchor=north west] at (0, 3.5) {$P^\oplus$};

    \draw (0,4) rectangle ++(5,4) coordinate[midway] (m);
    \draw [fill=TEETrusted] (5,4) rectangle ++(5,4) coordinate[midway] (e) {};
    \draw (e) circle (1em);
    \draw (-1, 5.5) node {PS};

    \draw[-latex] (e) ++ (0, -1) -- (e |- d.north) coordinate (x);

\end{tikzpicture}
		\caption{Legitimate system, with only one channel for transmitting image data from the \ac{PS} to the \ac{PL}.}
		\label{fig:pplus}
	\end{subfigure}
	\hfill
	\begin{subfigure}[t]{0.48\linewidth}
		\centering
		\begin{tikzpicture}
    \fill [fill=TEETrusted] (5,3.5) rectangle ++(5,0.5);

    \draw (0,-0.5) rectangle ++(10, 4) coordinate[midway] (c) {};
    \draw (-1,1) node {PL};
    \node [draw, rectangle, minimum width=6em, minimum height=2em] at (5, 1) (d) {Accelerator};
    \node [anchor=north west] at (0, 3.5) {$P^\ominus$};

    \draw (0,4) rectangle ++(5,4) coordinate[midway] (m);
    \draw [fill=TEETrusted] (5,4) rectangle ++(5,4) coordinate[midway] (e) {};
    \draw (e) circle (1em);
    \draw (-1, 5.5) node {PS};

    \draw[-latex] (e) ++ (0, -1) -- (e |- d.north) coordinate (x);

    \draw (m) circle (1em) ++ (0, -1) coordinate (m);
    \draw [red,thick] [*-latex] (x) ++ (0.3, 0.8) -| (m);
\end{tikzpicture}
		\caption{Malicious version, with an additional channel duplicating secret data back to normal world.}
		\label{fig:pminus}
	\end{subfigure}
	\caption{The architecture of the trusted 2D convolution filtering application.}
	\label{fig:conv2dattack}
\end{figure}

For clarity, we illustrate a hypothetical attack scenario on the system without our design, and demonstrate how our design can prevent such an attack.
The attack aims at exfiltrating image data processed by the accelerator, which is illustrated in \autoref{fig:conv2dattack}.
For clarity, we designate the legitimate channel transmitting image data from the \ac{PS} to the \ac{PL} as $C^\oplus$.
Conversely, the malicious channel, duplicating image data and redirecting it back to the \ac{PS}, is denoted as $C^\ominus$.
The legitimate version of the \ac{FPGA} bitstream, containing only $C^\oplus$, is represented as $P^\oplus$ (\autoref{fig:pplus}).
In contrast, the malicious version of the \ac{FPGA} bitstream, incorporating both $C^\oplus$ and $C^\ominus$, is labelled as $P^\ominus$ (\autoref{fig:pminus}).

In the absence of our design, distinguishing between the two versions of the \ac{FPGA} bitstream is challenging for a remote user.
The remote user, having knowledge only of $C^\oplus$, may deploy a user-space application to process image data through this channel.
Simultaneously, a malicious cloud service provider has the capability to deploy the malicious version, $P^\ominus$, while falsely claiming to deploy the legitimate version, $P^\oplus$.
The malicious cloud service provider can execute a separate program that listens to the $C^\ominus$ channel, effectively exfiltrating the image data without the remote user's awareness or consent.
This highlights the vulnerability in the absence of our design, where malicious actors can exploit the lack of authentication and verification mechanisms.

In contrast, our design empowers the remote user to leverage the attestation mechanism for verifying the authenticity of the \ac{FPGA} bitstream.
The remote user must possess prior knowledge of the hash of the legitimate $P^\oplus$ and the related public key.
Upon receiving the attestation report, the remote user verifies it by examining the signature and the hash of the \ac{FPGA} bitstream.
If the malicious cloud service provider attempts to deploy $P^\ominus$, the remote user can detect the attack by comparing the hash in the attestation report with the hash of $P^\oplus$.
A discrepancy in the hashes prompts the remote user to reject the attestation report and terminate the session, thereby fortifying the system against unauthorized alterations to the \ac{FPGA} bitstream.

From the device's view, no prior knowledge of the application is required until deployment.
This grants better flexibility to the user and developer, enabling the deployment of various applications on the same board without the need for reprovisioning of the cryptographic engine.
This level of flexibility surpasses the approach proposed in \cite{ccs/EisenbarthGPSSW07}.
It also outperforms \cite{ccs/OhAPLP20}, which supports only storage-oriented applications.
Moreover, both the attestation procedure and subsequent application execution involve only the remote user and the device to be attested.
This stands in contrast to the solution proposed by \cite{IEEEcloud/XuSS14}, which requires the involvement of a third-party in the procedure.
Given that the CPU processors (\ac{PS}) and the FPGA (\ac{PL}) are integrated on a single chip, the link between them is not exposed to the outside world.
This prevents adversary from tampering with the communication channel between the \ac{PS} and the \ac{PL}, eliminating the need for an encrypted channel between them.
This allows faster communication between the \ac{PS} and the \ac{PL} compared to the solution proposed in \cite{asplos/Zhao0K22}.

\subsection{Performance Evaluation}

\begin{table}
	\centering
	\caption{Performance comparison between experimental configurations.}
	\label{tab:perf}
	\begin{tabular}{c|cc}
		\toprule
		{\bf Configuration} & {\bf Boot Time} & {\bf Throughput} \\
		\midrule
		Without Attestation & 2125.80 ms      & 28.80 MB/s       \\
		With Attestation    & 2167.45 ms      & 28.97 MB/s       \\
		\bottomrule
	\end{tabular}
\end{table}

We measured the system boot time for both system images and recorded the throughput of the accelerator reported by the application after booting.
The overhead caused by rerouting secret data to the TrustZone is not included in our evaluation, as it is always required regardless of extending the TEE to the FPGA and is not an overhead introduced by our work.
The results are presented in \autoref{tab:perf}.
The application achieves a throughput of approximately 28.80~MB/s without attestation and 28.97~MB/s with attestation.
The difference in throughput is negligible, well within the margin of error.
Boot time for the system is 2125.80~ms without attestation and 2167.45~ms with attestation, indicating that enabling attestation introduces an overhead of 41.65~ms, which is less than 2\% of the boot time and considered acceptable.
These results demonstrate that our proposed attestation scheme has a low overhead during boot time and no impact on the application's runtime throughput.

Since our security design is based on ARM TrustZone and the FPGA fabric functions solely as a computation component, there is no usability overhead aside from the implementation of the application accelerator itself.
Our design does not limit the complexity of the accelerators; the potential complexity depends on the resources available on a specific SoC platform, such as LUTs and BRAM.

\section{Discussion}\label{sec:discussion}

\subsection{Key Revocation}

Assuming that the root-of-trust is not compromised, the attestation key
generated with respect to a vulnerable or deprecated \ac{FSBL} or \ac{ATF} can
be easily updated by loading a new version of the FSBL or ATF into the system.
This key derivation protocol ensures that a new pair of attestation keys will be
generated for the new version of the FSBL or ATF.
Additionally, the provisioning service can publish a revocation list of the
vulnerable or deprecated versions to prevent adversaries from loading them.
The service can also revoke the attestation key by publishing a certificate
revocation list.

However, if the root-of-trust is compromised, the feasibility of key revocation
depends on the nature of the hardware device.
If the device root-of-trust is backed by eFUSE and no margin is designed on the
eFUSE, revocation would be infeasible.
On the other hand, if the device root-of-trust is backed by a battery-backed
RAM, revocation would be possible by erasing the battery-backed RAM and
rewriting a new device secret.
This process will need to be performed by the provisioning service and may also
be completed remotely if encrypted key provisioning is available on the hardware
device.

\subsection{Cold-Boot Attack}

A cold-boot attack \cite{cold-boot-attack} exploits the fact that data stored in
RAM can persist for a brief period even after the power has been turned off.
By rapidly cooling the RAM chips, an attacker can preserve the data in the RAM
for a longer period and then remove the chips to read the data using specialized
tools.
This can allow an attacker to retrieve sensitive information such as encryption
keys or passwords that were stored in the RAM before the device was turned off.
Cold-boot attacks can bypass security measures such as full-disk encryption and
are particularly effective against devices that are not powered off regularly,
such as servers or embedded systems.

Standard TrustZone architecture does not provide any protection against
cold-boot attacks due to the lack of memory encryption.
However, there are two possible methods to implement memory encryption: hardware
and software.
A hardware-based approach involves adding an encryption engine to the memory
controller so that any data sent to external memory is encrypted.
On the other hand, software-based solutions to memory encryption are available,
such as those provided in \cite{trustcom/Horsch0W17} and \cite{dac/HuaUZS22}.

If memory encryption is not available, another mitigation technique is to use
\ac{OCM} as the main memory and external DDR as secondary memory.
With a paging mechanism, when the OCM is running out of space, it can kick out
some pages, encrypt them, and dump them to external DDR.
This approach ensures that the data stored in the external DDR is always
encrypted, thus reducing the risk of cold-boot attacks.

However, while cold-boot attacks can successfully exfiltrate data stored in
memory, including application data and attestation keys, the effects of the
attack are typically limited to the single application on the single device.
This isolation ensures that the operation of other applications or devices is
not affected by the attack.

\section{Related Work}\label{sec:related-works}

\paragraph{Rack-scale Solution}
Zhu et al. proposed a Platform as a Service (PaaS) solution called HETEE \cite{sp/ZhuH0WCZWZYZM20}.
It is a rack-scale server that incorporates off-the-shelf hardware accelerators such as GPUs and FPGAs.
Isolation and attestation are achieved through a security controller implemented on a PCIe ExpressFabric.
HETEE is physically protected within a tamper-resistant chassis that includes intrusion detection and response mechanisms.
However, there is a backdoor to open the chassis for maintenance purposes without triggering the protection mechanism.
This backdoor could provide a cloud service provider with an opportunity to misbehave and compromise the system.

There are also CISCO NCS 540 series routers \cite{cisco/ncs540}, which provide a trustworthy system with secure boot and attestation features.
They rely on the TPM chip to provide a hardware root-of-trust and SELinux to ensure access control at runtime.
However, the commercial solution has its limitations:
(1) The system includes all the hardware, firmware, operating system, and runtime within the trust boundary as part of the \ac{TCB}.
The \ac{TCB} is too large and complex, which increases the attack surface.
(2) The system is closed-source due to its proprietary nature, making it harder to evaluate the security of the system.
(3) The system is built on multiple chips, which exposes the risk of physical attacks against the physical links connecting the chips.

\paragraph{GPU-oriented Solution}
In recent years, attempts have been made to support a GPU-based trusted execution environment by forcing hardware changes.
Graviton, proposed by Volos et al. \cite{osdi/VolosVB18}, enhanced the current GPU card by equipping the GPU internal command processor with TEE features.
This additional isolation prevents the device driver, running on the potentially malicious host machine, from directly accessing GPU internal states.
HIX \cite{asplos/JangTKSH19}, proposed by Jang et al., modified the CPU chip by adapting the GPU driver to run in an SGX enclave, which required a change to the CPU-GPU I/O interconnect.
While these proposals have their merits, they are not feasible at present due to hardware changes required.

Recent research has explored combining ARM TrustZone and GPUs to create trusted execution environments, as seen in StrongBox \cite{ccs/DengWYLNLLYHCZ22} and Cronus \cite{micro/JiangQSCZWCZLC22}.
StrongBox utilizes an existing ARM GPU hardware and assumes exclusive access to the secure world for the trusted application, while Cronus allows for shared resources between the GPU and secure world.
However, there remains a need for a trusted heterogeneous computing platform that incorporates FPGAs to handle tasks that FPGAs excel at, such as network packet processing in data center environments or other embedded applications.
This platform should provide both security and deployment flexibility to meet the needs of diverse computing environments.

CAGE~\cite{University2024CAGECA} and ACAI~\cite{SridharaBSKAS24} are recent works that integrate GPUs into the trust domain established by Arm Confidential Compute Architecture (CCA).
These designs focus on cloud computing, where the ARM processor is virtualized, and hardware resources are shared among multiple VM tenants.
In contrast, our design targets edge computing, where the ARM processor is dedicated to a single user, and the FPGA is used for acceleration.
By adhering to a single-tenant model, we eliminate potential side-channel attacks that may occur due to shared resources.
Although limiting the number of VMs on a node to one may avoid resource sharing, the overall infrastructure, with the hypervisor and CCA as their main highlight, will be too heavy compared to our design.

\paragraph{Trusted Computing using FPGA}
Some works have attempted to incorporate FPGAs in innovative TEE designs, but often rely on using FPGAs as emulators for evaluation.
For example, Iso-X \cite{micro/EvtyushkinEOPGR14} is based on OpenRISC, while HECTOR-V \cite{asiaccs/NasahlSWM21} is built on RISC-V.
These works only build proof-of-concept TEEs on FPGA platforms instead of transforming FPGAs into trusted computation modules.
Split-Trust~\cite{mobisys/YaoTCSA23} proposed a hardware design that splits the hardware into multiple physically isolated components, where an FPGA is used to prototype the missing part of the hardware.
While calculating hashes using a TPM is a practical alternative, as proposed by Split-Trust, we chose to minimize dependency on additional hardware to keep the design simple, immediately deployable, and cost-effective.

Other works have attempted to provide security assurances based solely on FPGAs.
For instance, Thomas Eisenbarth et al. proposed introducing a soft TPM core into FPGA programs as part of the bitstream to build the chain of trust \cite{ccs/EisenbarthGPSSW07}.
Ken Eguro et al. claimed that FPGAs run as autonomous compute elements \cite{fpl/EguroV12}, so clients can offload sensitive parts of computation to FPGA, and they built a generic bootstrap program inside FPGA to receive and load user applications.
Hyunyoung Oh et al. proposed TRUSTORE \cite{ccs/OhAPLP20} as an extension of Intel SGX, which serves as a secure external storage device of SGX applications to help hide access patterns.
However, their attestation relies on a private key burned into the hardware during manufacturing, and FPGA platforms typically do not have secure entropy sources accessible, which requires specific hardware support.
Their following work, MeetGo \cite{access/OhNJCP21}, extends TRUSTORE to support general-purpose applications but retains similar limitations.
ShEF \cite{asplos/Zhao0K22} introduces another pure-FPGA solution that uses a shell for attestation and secure communication.
However, relying on a shell consumes additional FPGA resources and may reduce usability.
Furthermore, setting the boundary around the FPGA fabric introduces communication overhead between the FPGA and CPU due to channel encryption and decryption.
In contrast, our design, which targets heterogeneous computation, offers potentially better and faster solutions for tasks that require close collaboration between the CPU and FPGA.

Lei Xu et al. presented PFC \cite{IEEEcloud/XuSS14}, which proposed a privacy-preserving protocol to provide user data for FPGAs.
It is based on a proxy re-encryption scheme and makes use of the bitstream encryption scheme, which is typically used for intellectual property protection.
However, PFC requires the existence of a third-party proxy online, introducing communication overhead at runtime and making the system more complex.

\section{Conclusion}\label{sec:conclusion}

This paper presents a system that combines FPGA and TrustZone technology to enable secure heterogeneous computation.
The proposed system provides device and program authentication, establishes secure communication channels, and enables trusted computation of sensitive data.
The feasibility and effectiveness of the proposed solution were demonstrated through a prototype implementation on the Xilinx ZCU106 development board.
Furthermore, our protocol is formally verified using an automatic protocol verifier, providing additional confidence in its security properties.

In conclusion, our proposed solution provides a practical and effective approach to address the security and privacy challenges faced in FPGA-based heterogeneous computation.
It establishes a reliable foundation for deploying complex and sensitive applications on heterogeneous computing platforms.
Importantly, the protocol described in this paper is applicable to any FPGA IP, given the availability of ARM TrustZone.
We expect it to address security concerns as more FPGAs are deployed in the cloud, or via the cloud, with one or more third parties involved.
The integration of FPGA and TrustZone technology in our system model sets the stage for future research and advancements in this field.

\bibliographystyle{IEEEtran}
\bibliography{main}

\end{document}